\documentclass[aps,showpacs,amsmath,amssymb]{revtex4}
\usepackage{epsf,graphicx}
\usepackage{supercite}
\def\Bf#1{\mbox{\boldmath $#1$}}
\begin{document}
\title{Bose-Einstein Condensation Temperature of Dipolar Gas\\ 
in Anisotropic Harmonic Trap} 
\author{K. Glaum}
\email{glaum@physik.fu-berlin.de}
\affiliation{Fachbereich Physik, Freie Universit{\"a}t Berlin, 
Arnimallee 14, 14195 Berlin, Germany}
\author{A. Pelster}
\email{axel.pelster@uni-due.de}
\affiliation{Fachbereich Physik, Campus Duisburg, Universit{\"a}t Duisburg-Essen, 
Lotharstrasse 1, 47048 Duisburg, Germany}
\date{\today}
\begin{abstract}
We consider a dilute gas of dipole moments in an arbitrary harmonic trap and
treat both the short-range, isotropic delta-interaction and the long-range,
anisotropic dipole-dipole interaction perturbatively.
With this we calculate the leading shift of the critical temperature with respect to that of  
an ideal gas as a function of the relative orientation of the
dipole moments with respect to the harmonic trap axes.
In particular, we determine those magic angles, where the dipolar shift
of the Bose-Einstein condensation temperature vanishes. Furthermore,
we show for the parameters
of the ongoing  ${}^{52}$Cr-experiment in Stuttgart that this dipolar shift
can be enhanced by increasing the number of particles, the geometrical mean
trap frequency, and the anisotropy of the trap.
\end{abstract}
\maketitle
\section{Introduction}
Ultracold atomic quantum gases are many-body systems where a wide variety of
macroscopic quantum phenomena is observable 
\cite{Cornell,Ketterle1,Greiner1}. For the original Bose-Einstein condensates
(BECs) of alkali atoms, it has been sufficient to describe
the dominant two-particle interaction by a short-range and isotropic contact 
potential
\begin{eqnarray}
\label{CINT}
V^{({\rm int})}_{\delta} ( {\bf x} - {\bf x'}) = \frac{4 \pi \hbar^2 a_s}{M}
\, \delta ( {\bf x} - {\bf x'}) \, ,
\end{eqnarray} 
where $M$ stands for the mass of the particles.
The s-wave scattering length $a_s$ is tuned by using a so-called Feshbach 
resonance \cite{Ketterle2}, where
it can be varied over a broad range via an external magnetic field.
Recently, a BEC has also been realized in a dipolar quantum gas of ${}^{52}$Cr 
atoms \cite{Pfau6} where the magnetic dipole moments 
are around six times larger than those 
of alkali atoms. Therefore, the physical properties of such a chromium BEC
also depend on a long-range and anisotropic magnetic dipole-dipole interaction.
Other many-body systems with dipolar interactions are, for instance,
Rydberg atoms \cite{Rydberg1,Rydberg2} or atomic 
condensates where a strong electric field induces electric dipole
moments of the order of $10^{-2}$ Debye \cite{Yi1}.
Permanent dipole moments in heteronuclear 
molecules are much larger with typical values of 1 Debye, so their dipolar
effects could be a few hundred times stronger than those of chromium atoms 
\cite{Goral,Martikainen}. Such a gas of ultracold heteronuclear molecules is 
produced either by sophisticated cooling and trapping techniques 
\cite{Weinstein,Doyle,Meijer,Meerakker} or by photoassociation 
\cite{Sage,Wang,Mancini,Sengstock}. For all those systems
the specific aniosotropic dipole-dipole interaction reads
\begin{eqnarray}
\label{MINT}
V^{({\rm int})}_{DD} ({\bf x} - {\bf x'}) = - \frac{\mu_0}{4 \pi}\, 
\left\{ \frac{3 \left[ {\bf m}\,( {\bf x}-{\bf x'})\right]^2}
{|{\bf x}-{\bf x'}|^5} - \frac{{\bf m}^2}{|{\bf x}-{\bf x'}|^3} \right\}
+ \, \frac{\mu_0 m^2}{3} \, \delta ({\bf x} - {\bf x'})
\, ,
\end{eqnarray} 
where ${\bf m}$ denotes the magnetic (electric) dipole moment and
$\mu_0$ stands for the permeability constant (the reciprocal permittivity
constant). 
Note that this dipole-dipole interaction contains, apart from the usual first 
term, an additional contact term, which renormalizes the divergence at the 
origin. The latter contribution has been 
overlooked in most textbooks of electrodynamics, but is taken into account 
in more accurate representations as, for instance, in Ref.~\cite{Jackson}.
The dipolar interaction strength can be tuned for induced dipole moments
by varying the field strength and for permanent dipoles by 
using rotating magnetic (electric) fields \cite{Pfau2}. 
Combining this rotation technique with Feshbach resonances, 
will allow in the near future experiments where the
interaction varies from purely contact to purely dipolar.
The anisotropic dipole-dipole interaction gives rise to 
new condensate properties as, for instance, a characteristic anisotropic 
deformation of the expanding BEC which has recently been resolved 
experimentally in Ref.~\cite{Pfau7}. Also other
interesting dipolar phenomena are accessible experimentally as,
for instance, the occurrence of a Maxon-Roton in the excitation of dipolar 
BEC \cite{Santos1} or the instability of the ground state of dipolar BECs 
\cite{Rydberg1,Santos3,Eberlein1,Eberlein2}.\\

\begin{figure}[t]
\begin{center}
\includegraphics[scale=0.6]{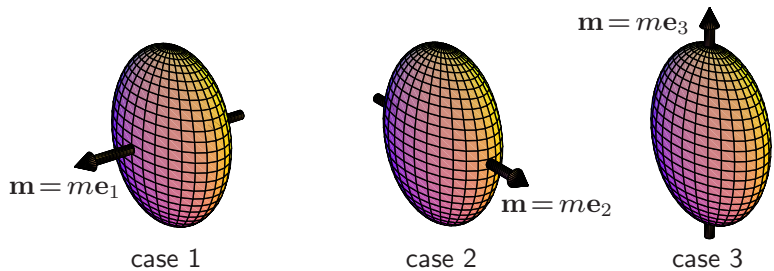} \hspace*{5mm}
\end{center}
\caption{\label{CASES} Axes of magnetic dipole-dipole interaction (\ref{MINT})
and harmonic trap (\ref{HT}) with frequencies $\omega_1 > \omega_2 > \omega_3$.
}
\end{figure}

In the present article we investigate how the critical temperature of a 
BEC is shifted due to the dipole-dipole interaction (\ref{MINT}). To 
this end we consider a dilute gas trapped in an arbitrary harmonic potential 
\begin{eqnarray}
\label{HT}
V({\bf x} ) = \frac{M}{2}\sum_{j=1}^3 \omega_j^2 x_j^2 \, ,
\end{eqnarray}
where $\omega_1, \, \omega_2, \, \omega_3$ denote the respective trap 
frequencies. Furthermore, we assume that, due to an additional external field, the dipole
moments ${\bf m}$ of all constituents are oriented along one axis, i.e. 
they are uniformly described by 
${\bf m}= m \,(\sin \alpha \cos \phi, \sin \alpha \sin \phi, \cos \alpha)$. 
We expect that the critical temperature depends crucially on the spherical angles 
$\alpha$ and $\phi$ which characterize the orientation of the dipole
moments with respect to the harmonic trap axes. The temperature shift should 
have local extrema when the dipole moments
are parallel to one of the harmonic trap axes as is illustrated in
Figure~\ref{CASES}. For instance, if the frequencies are ordered according to 
$\omega_1 > \omega_2 > \omega_3$, this leads to an 
additional interaction which is attractive along the 3-axis, repulsive along 
the 1-axis and can be both attractive or repulsive along the 2-axis.
Therefore, the particle density in the harmonic trap is increased (decreased)
in case~3 (case~1) in comparison with a pure contact interaction. Thus, 
we expect a resulting positive (negative) shift of the critical temperature 
due to the magnetic dipole-dipole interaction.
Furthermore, we follow Ref.~\cite{Glaum}
and suggest to determine differences of the critical temperatures
in the three configurations, as then the influence of the isotropic contact 
interaction cancels. Consequently, those differences represent a clear signal 
of the anisotropic dipole-dipole interaction.\\

We start our detailed analysis by briefly reviewing the calculation of the 
Bose-Einstein
condensation temperature for a dilute ideal Bose gas in Section~\ref{FRR}. 
Subsequently, we treat the influence of both interactions (\ref{CINT}) and (\ref{MINT}) 
in lowest order perturbatively. This is justified as the confinement of the harmonic trapping potential 
(\ref{HT}) removes critical long-wavelength fluctuations and reduces the fraction of
atoms taking part in nonperturbative physics at the transition point \cite{Boris}.
Therefore, we determine in Section~\ref{GFE} within first-order perturbation 
theory how the grand-canonical free energy and, thus, the particle number equation 
change under the influence of an arbitrary two-particle interaction potential. 
Additionally, we calculate in lowest perturbative order the location of the critical point
where the phase transition from a Bose gas to a BEC occurs. To this end we extract the
critical chemical potential from the first-order contribution to the self-energy.
This procedure is, in principle, self-consistent and corresponds to the
Hartree-Fock mean-field theory. However, we determine the critical chemical potential only
in the leading order and use it to obtain the critical temperture shift. Note that this provides a reliable
prediction since it is only linear in the interaction strength. Due to the Ginzburg criterion critical
fluctuations in the vicinity of the critial point violate the mean-field result in higher than first
order of the interaction strength \cite{Giorgini}.
In Section~\ref{MII} we specialize our result to the model interaction
potential (\ref{CINT}) and (\ref{MINT}) and combine it in Section~\ref{CT}
in order to derive the leading shift of the critical temperature. Furthermore,
we show how it is possible to tune the strength of this dipolar shift.
To this end  we investigate in Section~\ref{MANG} the magic angles where
the dipolar effects vanish. Finally, 
we discuss in Section~\ref{SEES} our theoretical findings for the
special parameters of the ongoing chromium experiment
in Stuttgart. In particular, we estimate which experimental parameters allow to
enhance the differences of the critical temperatures for two of the three 
configurations of Figure~\ref{CASES}.
\section{Interaction-free case}\label{FRR}
The quantum statistical properties of a dilute ideal Bose gas are 
determined by the grand-canonical partition function which follows from 
the functional integral
\begin{eqnarray}
\label{PF}
{\cal Z}^{(0)} = \oint {\cal D} \psi^* \, {\cal D} \psi \,\, 
e^{- {\cal A}^{(0)} [\psi^*,\psi] / \hbar} \, .
\end{eqnarray}
Here $\oint {\cal D} \psi^*  {\cal D} \psi$ sums over all bosonical field 
configurations $\psi({\bf x} , \tau )$ and $\psi^* ({\bf x} , \tau )$ which 
are periodic in imaginary time $\tau$.
Furthermore, the undisturbed euclidean action in (\ref{PF}) is
\begin{eqnarray}
\label{A0}
{\cal A}^{(0)} [ \psi^* , \psi ] = \hbar \, \int_0^{\hbar \beta} d\tau 
\int_0^{\hbar \beta} d\tau' \int d^3 x  \int d^3 x' 
\,\psi^*({\bf x},\tau)\, G^{(0)\,-1} ( {\bf x},\tau; {\bf x'},\tau' ) \,
\psi ({\bf x'},\tau') 
\end{eqnarray}
with the integral kernel
\begin{eqnarray}
G^{(0)\,-1} ( {\bf x},\tau; {\bf x'},\tau' ) = \frac{1}{\hbar}\, \delta( {\bf x} - {\bf x'} ) 
\delta( \tau - \tau' ) \,\left\{ \hbar \frac{\partial}{\partial \tau} -\frac{\hbar^2}{2M}\, {\Bf\Delta} +V({\bf x})
-\mu \,\right\}   \, ,
\label{SEE}
\end{eqnarray}
where $V({\bf x})$ is the external trap potential 
and $\mu$ stands for the chemical potential. 
The grand-canonical free energy of such an ideal Bose gas is given by
${\cal F}^{(0)} = - (\ln {\cal Z}^{(0)})/ \beta$
and may be evaluated from (\ref{PF}) as 
${\cal F}^{(0)} =  \mbox{Tr}\,\ln\,G^{(0)-1} / \beta$. 
For the harmonic trap potential (\ref{HT}), usually the condition 
$\hbar \beta \tilde{\omega} \ll 1$
holds, where $\tilde{\omega} = (\omega_1 \, \omega_2 \, \omega_3)^{1/3}$ denotes 
the geometric mean of the trap frequencies. This allows to determine
the grand-canonical free energy within
a semiclassical treatment \cite{Hagen} 
\begin{eqnarray}
\label{FF0}
{\cal F}^{(0)} = - \frac{1}{\beta (\hbar \beta \tilde{\omega})^3}\, \zeta_{4} \left( e^{\beta \mu} \right) \, ,
\end{eqnarray}
where $\zeta_{a} (z) = \sum_{n=1}^{\infty} z^n/n^a $ represents the polylogarithmic function.
Now we are interested in thermodynamic properties which result 
from a fixed average particle number
$N = - \partial {\cal F}^{(0)}/ \partial \mu$: 
\begin{eqnarray}
\label{N0}
N = \frac{1}{(\hbar \beta \tilde{\omega})^3}\,\zeta_3 \left( e^{\beta \mu} 
\right)
\, . 
\end{eqnarray}
In particular, we are interested in calculating the critical temperature 
where a macroscopic occupation
of the ground state sets in. The critical point can be immediately read 
off from Eq.~(\ref{N0}) to be $\mu_c^{(0)}=0$,
since there the polylogarithmic function starts to diverge. However, 
this critical point, where the phase
transition from a Bose gas to a BEC occurs, also coincides with a 
divergence of the correlation function
according to the theory of critical phenomena \cite{Verena}. To this 
end we determine the correlation
function of an ideal Bose gas from the function integral
\begin{eqnarray}
\label{CFF}
G^{(0)} ({\bf x} , \tau ; {\bf x}',\tau') = \frac{ 1 }{ {\cal Z}^{(0)} }
\oint {\cal D} \psi^* \, {\cal D} \psi \,\, \psi({\bf x} , \tau )
\psi^* ({\bf x}',\tau')\,e^{- {\cal A}^{(0)} [\psi^*,\psi] / \hbar} 
\, .
\end{eqnarray}
The semiclassical result reads
\begin{eqnarray}
\hspace{-0.8cm}G^{(0)} ({\bf x}, \tau ; {\bf x}',\tau') &=& \int 
\frac{d^3 p}{(2 \pi \hbar)^3} \,
e^{i \,{\bf p} ({\bf x}-{\bf x}')/ \hbar} \, \bigg\{ \Theta (\tau - \tau') 
\sum_{n=0}^{\infty} \, e^{- \frac{1}{\hbar} \left[
\frac{{\bf p}^2}{2 M} + V\left(\frac{{\bf x}+{\bf x'}}{2} \right)-\mu 
\right] \left( \tau - \tau' \!+ n \hbar \beta \right)}
\nonumber \\*[1mm]
\hspace{-0.8cm}&& + \, \Theta (\tau' - \tau) \sum_{n=1}^{\infty} \,
e^{- \frac{1}{\hbar} \left[
\frac{{\bf p}^2}{2 M} + V\left(\frac{{\bf x}+{\bf x'}}{2} \right)-\mu 
\right] \left( \tau - \tau' \!+ n \hbar \beta \right)} \bigg\} \, .
\label{G}
\end{eqnarray}
The Fourier-Matsubara transform for such a semiclassical expression is given by
\begin{eqnarray}
\label{FOURTRAFO}
G^{(0)} ({\bf p} , \omega_m ; {\bf x}) = \int_0^{\hbar \beta} d \tau \,
e^{ i \omega_m \tau } \int d^D x' \, e^{ - i {\bf p}{\bf x}'/\hbar } \, 
G^{(0)} \left( {\bf x} + \frac{\bf x'}{2} , \tau ; {\bf x} - \frac{\bf x'}{2},
0 \right) \, ,
\end{eqnarray}
where $\omega_m=2 \pi m / \hbar \beta$ denotes the Matsubara frequencies. 
Inserting (\ref{G}) in (\ref{FOURTRAFO}) yields
\begin{eqnarray}
G^{(0)} ({\bf p} , \omega_m ; {\bf x}) = \, \frac{\hbar}{- i \hbar \omega_m + 
\frac{{\bf p}^2}{2 M} + V ( {\bf x} ) - \mu} \, ,
\end{eqnarray}
which reveals explicitly that (\ref{CFF}) represents the functional
inverse of the integral kernel (\ref{SEE}). At
the critical point, where the correlation function (\ref{CFF}) diverges, 
the integral kernel vanishes. Within the semiclassical approximation 
of the Fourier-Matsubara transform of (\ref{SEE}) we thus obtain 
\begin{eqnarray}
\label{CRIT}
G^{(0)-1} ({\bf p} , \omega_m ; {\bf x}) = \, \frac{1}{\hbar} \left\{
- i \hbar \omega_m + \frac{{\bf p}^2}{2 M} + V ( {\bf x} ) - \mu \right\}
= 0 \, .
\end{eqnarray}
This equation can only be fulfilled for vanishing momentum ${\bf p} = {\bf 0}$
and Matsubara frequency $\omega_m=0$ at the critical chemical potential
\begin{eqnarray}
\label{V3-0}
\mu^{(0)}_c = \min_{\bf x} V ( {\bf x} ) \, .
\end{eqnarray}
In case of the harmonic potential (\ref{HT}) the minimum occurs in the center
of the trap. Thus, we reproduce the critical chemical potential 
$\mu_c^{(0)}=0$, which corresponds to the semiclassical 
value of the ground-state energy. Inserting this into Eq.~(\ref{N0}), we obtain for the
transition temperature for an interaction-free Bose gas 
\begin{eqnarray}
\label{TC0}
T_c^{(0)} = \frac{\hbar \tilde{\omega}}{k_B}\,\left[ \frac{N}{\zeta(3)} 
\right]^{1/3} \, ,
\end{eqnarray}
where $\zeta (a) = \sum_{n=1}^\infty 1/ n^{a}$ is the so-called Riemann 
zeta function. Note that in more complicated traps, which arise, for instance,
when the Bose gas is rotated, the potential minimum could occur far away 
from the origin, thus resulting in a non-vanishing critical chemical potential
\cite{Kling}.\\

From the critical temperature (\ref{TC0}) we read off that 
$\hbar \beta_c^{(0)} \tilde{\omega} \sim N^{-1/3}$, i.e.~the condition for the semiclassical
treatment is fulfilled within the thermodynamic limit $N \to \infty$. For a finite
number of particles $N$, it is necessary to go beyond the semiclassical 
approximation and to take into account quantum corrections in a systematic way  
\cite{Grossmann,Haugset,Dalfovo,Yukalov}. In lowest order this finite-size effect
in harmonic traps follows from identifying $\mu_c^{(0)}$ not with the semiclassical value
of the ground-state energy but with its quantum mechanical value $E_G= 3 \hbar \overline{\omega} / 2$,
where $\overline{\omega}= (\omega_1 + \omega_2 + \omega_3)/3$ denotes the arithmetic mean of the trap
frequencies. Evaluating the polylogarithmic function in (\ref{N0}) perturbatively
up to the leading order, we obtain for the critical temperature the finite-size effect
\begin{eqnarray}
\label{FS}
\left( \frac{\Delta T_c}{T_c^{(0)}}\right)_{FS} = \,
- \frac{\zeta(2) \overline{\omega}}{2 \zeta^{2/3}(3) \tilde{\omega} N^{1/3}} 
\, .
\end{eqnarray}
The reduction of the critical temperature of an ideal gas decreases with increasing
the particle number $N$.
\section{First-order Perturbation theory}\label{GFE}
In this section we investigate the influence of an arbitrary weak 2-particle interaction 
upon the thermodynamic properties of dilute Bose gases 
within Feynman's diagrammatic technique of
many-body theory \cite{Huang,Abrikosov,Fetter,Negele,Gross,Mahan}. 
\subsection{Feynman Rules}\label{FEYN}
We start with
defining the grand-canonical partition function of the full problem 
\begin{eqnarray}
\label{PFful}
{\cal Z} = \oint {\cal D} \psi^* \, {\cal D} \psi \,\, 
e^{- \left( {\cal A}^{(0)} [\psi^*,\psi] + {\cal A}^{({\rm int})}[\psi^*,\psi]
\right) / \hbar} 
\end{eqnarray}
and the associated correlation function
\begin{eqnarray}
\label{CFFful}
G ({\bf x} , \tau ; {\bf x}',\tau')= \frac{1}{\cal Z}
\oint {\cal D} \psi^* \, {\cal D} \psi \,\, \psi({\bf x} , \tau )
\psi^* ({\bf x}',\tau')\,e^{- \left( {\cal A}^{(0)} [\psi^*,\psi] 
+ {\cal A}^{({\rm int})}[\psi^*,\psi] \right) / \hbar} 
\, .
\end{eqnarray}
In addition to
the interaction-free contribution to the action (\ref{A0}) we have also to take into account the 
interaction contribution. Due to the diluteness of the Bose gas we can restrict ourselves to the effects of  
a two-particle interaction $V^{({\rm int})} ({\bf x}- {\bf x}')$:
\begin{eqnarray}
\label{INT}
{\cal A}^{({\rm int})}[\psi^*,\psi] = \frac{1}{2} \int_0^{\hbar \beta} d\tau \int d^3 x \int d^3 x'\,\,
V^{({\rm int})} ({\bf x}- {\bf x}') \,
\psi^* ( {\bf x} , \tau ) \psi^* ( {\bf x}' , \tau ) \psi ( {\bf x} , \tau )\psi ( {\bf x}' , \tau )
\, .
\end{eqnarray}
By expanding the functional interals (\ref{PFful}), (\ref{CFFful}) in powers 
of $V^{({\rm int})} ({\bf x}- {\bf x}')$, we obtain expressions
which consist of 
interaction-free expectation values. These
are evaluated with the help of Wick's rule as a sum of Feynman integrals, 
which are pictured as diagrams constructed from
lines and vertices. A straight line with an arrow represents the 
interaction-free correlation function (\ref{CFF}):
\begin{eqnarray}
\label{PRO}
&&\hspace*{-6mm}
\includegraphics[scale=1.0]{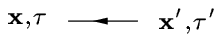}
\hspace*{2mm}
\equiv \hspace*{3mm} G^{(0)}({\bf x}, \tau ; {\bf x'},\tau')  
\, .
\end{eqnarray}
Furthermore, spatio-temporal integrals over the two-particle interaction 
potential are pictured by two vertices connected by a dashed line 
\begin{eqnarray}
\label{VE}
\raisebox{-1.5mm}{\includegraphics[scale=1.0]{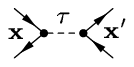}}
\hspace*{2mm} \equiv \hspace*{2mm} - \frac{1}{\hbar} \int_0^{\hbar \beta} d \tau \, \int d^3 x \,\int d^3 x'\,
V^{({\rm int})}({\bf x}-{\bf x'}) \hspace*{2mm}   \, .
\end{eqnarray}
In the following we apply these Feynman rules in order to 
determine the grand-canonical free energy and the self-energy
within first-order perturbation theory for the harmonic trap 
potential (\ref{HT}) and
an arbitrary interaction potential $V^{({\rm int})}({\bf x}-{\bf x'})$. 
\subsection{Grand-Canonical Free Energy}
Up to the first order in the two-particle interaction the grand-canonical
free energy ${\cal F} = - (\ln {\cal Z})/\beta$ reads with the diagrammatic
rules mentioned above
\begin{eqnarray}
\raisebox{-3.5mm}{\includegraphics[scale=.9]{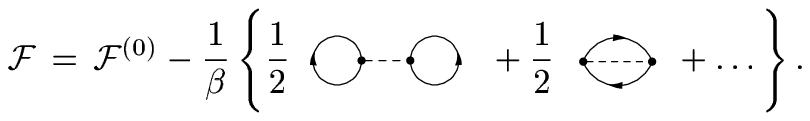}}
\label{F1}
\end{eqnarray}
By passing we note that those two and all higher-order connected vacuum 
diagrams follow together with their proper weights from solving a graphical 
recursion relation \cite{Devreese}.
The first term in (\ref{F1}) is the interaction-free contribution to the 
grand-canonical free energy (\ref{FF0}).
The second and third term in (\ref{F1}) are called the direct or Hartree-like
and exchange or Fock-like 
vacuum diagrams, respectively, which correspond to the 
following analytical expressions:
\begin{eqnarray}
\label{F1A}
{\cal F}^{({\rm D})}& = & \frac{1}{2 \hbar \beta}\,
\int_0^{\hbar \beta} d \tau \, \int d^3 x \, \int d^3 x' \,V^{({\rm int})}({\bf x}-{\bf x'}) \,
G^{(0)}({\bf x} , \tau ; {\bf x},\tau) G^{(0)}({\bf x}' , \tau ; {\bf x}',\tau) \, ,\\
\label{F1B}
{\cal F}^{({\rm E})} & = & \frac{1}{2\hbar \beta}\,
\int_0^{\hbar \beta} d \tau \, \int d^3 x \, \int d^3 x' \,V^{({\rm int})}({\bf x}-{\bf x'}) \,
G^{(0)}({\bf x} , \tau ; {\bf x}',\tau) G^{(0)}({\bf x}' , \tau ; {\bf x},\tau) \, .
\end{eqnarray}
Both contain the interaction-free correlation function (\ref{G}) 
with equal imaginary times. In order to guarantee the normal operator 
ordering within the functional integral formalism, 
this equal-time correlation function $G^{(0)}({\bf x} , \tau ; {\bf x}',\tau)$
must be interpreted as $G^{(0)}({\bf x} , \tau ; {\bf x}',\tau^+)$ \cite{Fetter,Negele}.
Here we have introduced $\tau^+$ as an imaginary time which is infinitesimally later than $\tau$.
With this and a Fourier transformation of the interaction potential 
we obtain for 
the harmonic trap potential (\ref{HT}):
\begin{eqnarray}
\label{FD}
{\cal F}^{({\rm D})} & = & \frac{1}{2 (\hbar \beta \tilde{\omega})^{6}} \sum_{n=1}^\infty \sum_{n'=1}^\infty
\frac{e^{(n+n')\beta \mu}}{n^3 {n'}^3}  \int \frac{d^3 q}{(2 \pi \hbar)^3}\,V^{({\rm int})} ( {\bf q} )
\,\exp \left\{ - \sum_{j=1}^3 \frac{(n+n') \,q_j^2}{2 \hbar^2 \beta M n n' \omega_j^2} \right\}\, , \\
\label{FE}
{\cal F}^{({\rm E})} & = &\frac{1}{2 (\hbar \beta \tilde{\omega})^3} \sum_{n=1}^\infty \sum_{n'=1}^\infty
\frac{e^{(n+n')\beta \mu}}{(n + n')^3}  \int \frac{d^3 q}{(2 \pi \hbar)^3}\,V^{({\rm int})} ( {\bf q} )
\,\exp \left\{ - \frac{\beta n n' {\bf q}^2}{2 M (n+n')} \right\} \, .
\end{eqnarray}
\subsection{Self-Energy}
In order to determine the location of the phase transition from a Bose gas 
to a BEC, we follow the physical notion elucidated in the ideal case of the 
previous section and investigate 
when the correlation function (\ref{CFFful}) diverges, i.e., when 
its functional inverse 
$G^{-1} ({\bf x} , \tau ; {\bf x}',\tau')$ vanishes.
Due to the two-particle interaction, it decomposes into the integral kernel (\ref{SEE}) and
the self-energy $\Sigma ({\bf x}, \tau ; {\bf x'},\tau')$ according to
\begin{eqnarray}
\label{SELF}
G^{-1} ({\bf x}, \tau ; {\bf x'},\tau') = G^{(0)\,-1}({\bf x}, \tau ; 
{\bf x'},\tau') - \Sigma ({\bf x}, \tau ; {\bf x'},\tau')  \, ,
\end{eqnarray}
which is equivalent to the Dyson equation \cite{Huang,Abrikosov,Fetter,Negele,Gross,Mahan}
\begin{eqnarray}
G ({\bf x}, \tau ; {\bf x'},\tau') &=& G^{(0)}({\bf x}, \tau ; {\bf x'},\tau') \nonumber \\
&&+
\int d^3 x'' \int d^3 x''' \int_0^{\hbar \beta} d \tau''  \int_0^{\hbar \beta} d \tau''' G ({\bf x}, \tau ; {\bf x''},\tau'')
 \Sigma ({\bf x''}, \tau'' ; {\bf x'''},\tau''')  G^{(0)}({\bf x'''}, \tau''' ; {\bf x'},\tau') \, .
\label{DY}
\end{eqnarray}
Therefore, a perturbative evaluation of the self-energy yields via (\ref{DY}) self-consistent 
interaction corrections to the free correlation function (\ref{CFF}). In lowest
perturbative order the self-energy is defined by the diagrams
\begin{eqnarray}
\hspace*{-1mm}\raisebox{-4.0mm}{\includegraphics[scale=1.0]{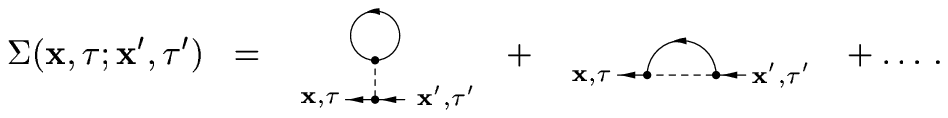}} \, ,
\label{sigma}
\end{eqnarray}
which correspond to the analytical expressions
\begin{eqnarray}
\Sigma^{({\rm D})}({\bf x} , \tau ; {\bf x}',\tau') &=& \frac{-1}{\hbar} \,
\delta( \tau - \tau' )  \delta ({\bf x} - {\bf x}' ) \int d^3 x'' \, 
V^{({\rm int})} ( {\bf x} - {\bf x}'' ) G^{(0)} ( {\bf x''}, \tau ;{\bf x''}, 
\tau ) \, , \label{SEDA}
\\
\Sigma^{({\rm E})}({\bf x} , \tau ; {\bf x'},\tau') &=& \frac{-1}{\hbar} \,
\delta( \tau - \tau' ) V^{({\rm int})} ( {\bf x} - {\bf x'} ) 
G^{(0)} ( {\bf x}, \tau ;{\bf x'}, \tau ) \, .
\label{SED}
\end{eqnarray}
Note that the diagrams  (\ref{sigma}) follow from amputating a line in the connected vacuum 
diagrams of the grand-canonical free energy \cite{CL1,CL2}. Thus, using the 
first-order perturbative approximation (\ref{sigma}) for the self-energy
corresponds via the Dyson equation (\ref{DY}) to a nonperturbative Hartree-Fock mean-field theory
for the correlation function.\\

To find the zero of (\ref{SELF}), we consider its Fourier-Matsubara
transform. Together with (\ref{SEE}) this leads to the equation
\begin{eqnarray}
\label{GDECO}
\frac{1}{\hbar} \left\{ - i \hbar \omega_m + 
\frac{{\bf p}^2}{2 M} + V ( {\bf x} ) - \mu \right\} - \Sigma ( {\bf p} , \omega_m ; {\bf x} ) = 0 \, ,
\end{eqnarray}
where the Fourier-Matsubara transform of the self-energy follows in analogy
to Eq.~(\ref{FOURTRAFO}) from
\begin{eqnarray}
\label{FMT}
\Sigma ( {\bf p} , \omega_m ; {\bf x} ) = 
\int_0^{\hbar \beta} \!\! d \tau \,
e^{ i \omega_m \tau } \! \int d^D x' \, e^{ - i {\bf p}{\bf x}'/\hbar } \, 
\Sigma \left( {\bf x} + \frac{\bf x'}{2} , \tau ; {\bf x} - \frac{\bf x'}{2},
0 \right) \, . 
\end{eqnarray}
Eq.~(\ref{GDECO}) is fulfilled up to the lowest perturbative order at 
vanishing momentum ${\bf p} = {\bf 0}$
and Matsubara frequency $\omega_m=0$ as in the previous section. 
Furthermore, we conclude that the additional contributions (\ref{SEDA}), 
(\ref{SED}) in (\ref{GDECO}) do not change the location 
${\bf x}_{\rm min} = {\bf 0}$ of the potential minimum. Therefore, the critical 
chemical potential is given up to first
order by $\mu_c = - \hbar \, \Sigma ( {\bf 0} , 0 ; {\bf 0} )$.
Identifying the direct and the exchange contribution with
${\displaystyle \mu_c^{(\rm D)} = - \hbar \, \Sigma^{(\rm D)}
( {\bf 0} , 0 ; {\bf 0} ) }$ and ${\displaystyle \mu_c^{(\rm E)} = - \hbar 
\, \Sigma^{(\rm E)}( {\bf 0} , 0 ; {\bf 0} ) }$,
we obtain 
${\displaystyle \mu_c = \mu_c^{({\rm D})}+  \mu_c^{({\rm E})}+ \ldots}$,
where the respective terms read according to (\ref{G}) 
\begin{eqnarray}
\label{MSD}
\mu_c^{({\rm D})}
& = & \frac{1}{(\hbar \beta \tilde{\omega})^3}\sum_{n=1}^\infty\frac{1}{n^3} 
\int \frac{d^3 q}{(2 \pi \hbar)^3}\,V^{({\rm int})} ( {\bf q} )
\,\exp \left\{ - \sum_{j=1}^3 \frac{q_j^2}{2 \hbar^2 \beta M n \omega_j^2} 
\right\}\, , \\*[2mm]
\label{MSE}
\mu_c^{({\rm E})} & = & \sum_{n=1}^\infty 
\int \frac{d^3 q}{(2 \pi \hbar)^3}\,V^{({\rm int})} ( {\bf q} ) \,
\exp \left\{ - \sum_{j=1}^3 \frac{\beta n \, q_j^2 }{ 2 M } \right\} .
\end{eqnarray}
\section{Model Interaction}\label{MII}
In this section we specialize our general formulas for the two-particle
interaction between dipolar bosons which
contains both the contact interaction (\ref{CINT}) and the 
dipole-dipole interaction (\ref{MINT}). At first, we briefly discuss whether
applying this model interaction is physically reasonable. Recently, it has been
suggested in Refs.~\cite{Bortolotti,Ronen} that 
the $s$-wave scattering length $a_s$ could strongly depend on the dipole moment.
However, there it has also been shown that for
dipolar interaction strengths, which are not larger than the s-wave scattering strength, 
the latter is only rescaled
by a moderate factor and remains positive. For the calculations 
in the present work we assume that this condition is fulfilled. 
Furthermore, we mention that the contact interaction 
(\ref{CINT}) represents an effective pseudopotential, see e.g.~Refs.~\cite{Yang,Huang}. 
An analogous pseudopotential for anisotropic interactions has recently 
been introduced in Refs.~\cite{Derevianko,You} which contains,
apart from the usual 
dipole-dipole interaction (\ref{MINT}) in the s-d scattering channel, also another 
part from the d-s scattering channel. The contribution of the latter turns out to be 
nonlocal in momentum space and is therefore difficult 
to handle. Here we assume that we are not in the vicinity of any dipolar shape resonance 
and, hence, adopt the anisotropic interaction in the original form 
(\ref{MINT}). Since there exists up to now no experimental evidence that a
scattering in the d-s channel is relevant,
our model seems to be valid for all current experimental situations. 
\subsection{Fourier Representation}
Now we combine both interactions (\ref{CINT}), (\ref{MINT}) and write 
our model interaction in the following form:
\begin{eqnarray}
\label{CM-INT}
V^{({\rm int})} ({\bf x} - {\bf x'}) \, = \, \frac{4 \pi \hbar^2 a}{M} \, 
\delta ({\bf x} - {\bf x'}) - \frac{\mu_0}{4 \pi}\, 
\left\{ \frac{3 \left[ {\bf m}\,( {\bf x}-{\bf x'})\right]^2}
{|{\bf x}-{\bf x'}|^5} - \frac{{\bf m}^2}{|{\bf x}-{\bf x'}|^3} \right\} \, .
\end{eqnarray} 
Here we used the effective scattering length 
\begin{eqnarray}
\label{a-as}
a \equiv a_s + \, \frac{\mu_0 m^2 M}{12 \pi \hbar^2} 
\end{eqnarray}
for describing the strength of the contact interaction.
Note that both terms in the effective scattering length 
(\ref{a-as}), i.e. the original s-wave scattering length $a_s$ and
the magnetic dipole contribution, are physically inseparable. Therefore, we have
to identify (\ref{a-as}) with the experimentally measurable scattering length of
atoms with magnetic dipole moments.  
In the following we will use the dimensionless parameter 
\begin{eqnarray}
\label{EDD}
\epsilon_{DD} = \frac{ \mu_0 m^2 M}{12 \pi \hbar^2 a} \, ,
\end{eqnarray}
which is a measure of the strength of the dipole-dipole interaction relative 
to the effective scattering energy. 
In the limit of a vanishing scattering length $a_s \to 0$ the model 
interaction (\ref{CM-INT}) describes the pure dipolar interaction 
(\ref{MINT}). This corresponds to the value $\epsilon_{DD} = 1$,
which plays a special role by investigating the stability of a dipolar BEC at zero
temperature within the Thomas-Fermi approximation \cite{Eberlein1,Eberlein2}.
Under certain conditions one expects at this dipole interaction strength that 
a cloud of dipolar particles collapses into a thin wire along the magnetization 
direction. In contrast, the value $\epsilon_{DD} \to \infty$ is only
achieved in the limit $a_s \to -\mu_0 m^2 M /12 \pi \hbar^2$ 
which does not correspond to any distinguished situation.
Furthermore, if we consider the limit of a vanishing dipole
moment $m \to 0$, the model interaction reduces 
to the pure contact interaction (\ref{CINT}).\\ 

Now we determine the Fourier representation of our model interaction (\ref{CM-INT}).
The Fourier transform of the contact interaction is simply a constant.
The Fourier transformation of the anisotropic term 
has to be evaluated with special care, as 
the corresponding Fourier integral is UV-divergent.
One possibility to regularize this singularity is to introduce an UV cut-off 
distance which physically takes into account that the atoms cannot overlap 
\cite{Goral}. Thus, the Fourier transform is determined for a finite UV 
cut-off distance which is, finally, allowed to go to zero. Using
spherical coordinates, where the dipole moment is described by
${\bf m}= m \,(\sin \alpha \cos \phi, \sin \alpha \sin \phi, \cos \alpha)$, we obtain
the following Fourier transform:
\begin{eqnarray}
\label{DI}
V^{({\rm int})} ( {\bf q} ) &=& \frac{4 \pi \hbar^2 a}{M} \, - \, 
\frac{\mu_0 m^2}{3} \, + \, \frac{\mu_0 m^2}{{\bf q}^2 } \, 
\Big[ \sin^2 \! \alpha \left( \cos^2 \! \phi \,\, q_1^2 + \sin^2 \! \phi \,\,
q_2^2 \right) + \cos^2 \! \alpha \,\, q_3^2  
\nonumber \\
& & + \, 2 \sin\alpha \cos\alpha \left( \cos \phi \, q_1 q_3 + \sin \phi \, 
q_2 q_3 \right) + 2 \sin^2 \! \alpha \sin \phi \cos \phi \, q_1 q_2 \Big] \, .
\end{eqnarray} 
Another approach for obtaining this result (\ref{DI}) has been explored 
in quantum electrodynamics in the context of calculating the transverse 
delta function with a regularization function \cite{Cohen}. Note that the result (\ref{DI})
also follows from applying the distributional identity \cite{Eberlein2}
\begin{eqnarray}
\frac{3 x_i x_j - \delta_{ij} {\bf x}^2}{| {\bf x}|^5}  \, = \,  
\frac{4 \pi}{3}\,\delta_{ij}\, \delta({\bf x}) +
\frac{\partial^2}{\partial x_i \partial x_j}\,\frac{1}{| {\bf x}|}  
\, ,
\end{eqnarray}
which can be regarded as defining the dipole-dipole interaction as second 
partial derivatives of the Coulomb potential.
\subsection{Evaluation of Integrals}
Now we evaluate the influence of the interaction (\ref{DI}) upon the 
thermodynamic quantities of interest. To this end we remark
that the remaining integrals (\ref{FD}), (\ref{FE})
and (\ref{MSD}), (\ref{MSE}) are of the form
\begin{eqnarray}
\label{MIint}
I^{({\rm int})} \left( \frac{a_1}{a_2},\frac{a_1}{a_3} \right) \!&=&\! - \, 
\frac{3}{\mu_0 m^2 a_1 a_2 a_3} \int_{- \infty}^\infty 
\frac{d q_1}{\sqrt{\pi}} \int_{- \infty}^\infty \frac{d q_2}{\sqrt{\pi}}  
\int_{- \infty}^\infty \frac{d q_3}{\sqrt{\pi}}
\,\, V^{({\rm int})} ({\bf q}) \,  \exp
\left( - \frac{q_1^2}{a_1^2} - \frac{q_2^2}{a_2^2}- \frac{q_3^2}{a_3^2} 
\right) \, .
\end{eqnarray}
In order to calculate this quantity, we define the integrals
\begin{eqnarray}
\label{MI1}
I^{(j)} \left( \frac{a_1}{a_2},\frac{a_1}{a_3} \right) \!&=& 
\frac{3}{a_1 a_2 a_3} \int_{- \infty}^\infty 
\frac{d q_1}{\sqrt{\pi}} \int_{- \infty}^\infty \frac{d q_2}{\sqrt{\pi}}  
\int_{- \infty}^\infty \frac{d q_3}{\sqrt{\pi}}
\, \left( \frac{1}{3} - \frac{1}{3 \, \epsilon_{DD}} - 
\frac{q_j^2}{{\bf q}^2} \right) \, \exp
\left( - \frac{q_1^2}{a_1^2} - \frac{q_2^2}{a_2^2}- \frac{q_3^2}{a_3^2} 
\right) \, ,
\end{eqnarray}
which occur for the configurations $j=1$, 2, and 3 of Figure \ref{CASES}, 
respectively. Due to the symmetry of its integrand, 
the full problem (\ref{MIint}) is then solved by
\begin{eqnarray}
\label{MI-II-III}
I^{({\rm int})} \left( \frac{a_1}{a_2},\frac{a_1}{a_3} \right) \,=\, 
\sin^2 \! \alpha \cos^2 \! \phi \,\, 
I^{(1)} \left( \frac{a_1}{a_2},\frac{a_1}{a_3} \right) + 
\sin^2 \! \alpha \sin^2 \! \phi \,\, 
I^{(2)} \left( \frac{a_1}{a_2},\frac{a_1}{a_3} \right) 
+ \cos^2 \! \alpha \,\, I^{(3)} \left( \frac{a_1}{a_2},\frac{a_1}{a_3} \right) 
\, .
\end{eqnarray}
From this representation we obtain immediately that 
$I^{({\rm int})}( a_1/a_2,a_1/a_3 )$ is extremal for arbitrary 
parameters $a_1$, $a_2$ and $a_3$ only if
\begin{eqnarray}
\label{ANGLES}
\left\{ \hspace*{3mm}
\begin{array}{@{}lll}
\alpha = \pi/2 \,  , \quad & \phi = 0, \pi: & 
\hspace*{0.5cm} 
{\displaystyle I^{({\rm int})} \left( \frac{a_1}{a_2},\frac{a_1}{a_3} \right) 
= I^{(1)} \left( \frac{a_1}{a_2},\frac{a_1}{a_3} \right)} \, ,
\\*[4mm]
\alpha = \pi/2 \, , \quad & \phi = \pi/2, 3 \pi/2: & 
\hspace*{0.5cm} 
{\displaystyle I^{({\rm int})} \left( \frac{a_1}{a_2},\frac{a_1}{a_3} \right) 
= I^{(2)} \left( \frac{a_1}{a_2},\frac{a_1}{a_3} \right)} \, ,
\\*[4mm]
\alpha = 0 \, ,  & \phi \,-\, {\rm arbitrary}: & 
\hspace*{0.5cm} 
{\displaystyle I^{({\rm int})} \left( \frac{a_1}{a_2},\frac{a_1}{a_3} \right) 
= I^{(2)} \left( \frac{a_1}{a_2},\frac{a_1}{a_3} \right)} \, .
\end{array} \right. 
\end{eqnarray}
Whether these extrema lead to local maxima or minima of $I^{({\rm int})}$
depends on the concrete values of the coefficients $I^{(1)}$, $I^{(2)}$ and 
$I^{(3)}$. To calculate them we mention the properties
\begin{eqnarray}
\label{ItoII}
I^{(2)} \left( \frac{a_1}{a_2},\frac{a_1}{a_3} \right) = 
I^{(1)} \left( \frac{a_2}{a_3},\frac{a_2}{a_1} \right) \, ,
\hspace{6mm} I^{(3)} \left( \frac{a_1}{a_2},\frac{a_1}{a_3} \right) = 
I^{(1)} \left( \frac{a_3}{a_1},\frac{a_3}{a_2} \right) \, ,
\end{eqnarray}
which simply follow from definition (\ref{MI1}). 
Consequently, we only need to determine one of the three integrals, for 
instance, that for $j=1$.
Applying the Schwinger proper-time representation 
${1/{\bf q}^2 = \int_0^\infty d \tau \, e^{ -{\bf q}^2 \tau }}$ \cite{Verena}, 
we arrive at the expression
\begin{eqnarray}
\label{F-Integral}
I^{(1)} \left( \frac{a_1}{a_2},\frac{a_1}{a_3} \right) = 1 - \, 
\frac{1}{\epsilon_{DD}} \, - \, \frac{3}{2 \, a_1 a_2 a_3} 
\int_0^{\infty} d \tau \, \frac{ 1 }{ \left( \tau + 1/a_1^2 \right)^{3/2} \, 
\left( \tau + 1/a_2^2 \right)^{1/2} \, \left( \tau + 1/a_3^2 \right)^{1/2} } 
\hspace*{2mm} .
\end{eqnarray}
With the help of substitutions the remaining integral is reduced to
incomplete elliptic functions of the first and second kind
\cite[(8.111)]{Gradshteyn}
\begin{eqnarray}
F( \phi, k ) = \int_0^{\sin \phi} \!\! d x \,\, \frac{1}
{ \left( 1-x^2 \right)^{1/2} \, \left( 1-k^2 x^2 \right)^{1/2} } 
\hspace*{7mm} , \hspace*{5mm} E( \phi, k ) = 
\int_0^{\sin \phi} \!\! d x \,\, \frac{ \left( 1-k^2 x^2 \right)^{1/2} }
{ \left( 1-x^2 \right)^{1/2} } \, .
\end{eqnarray}
We present the result in the following form:
\begin{eqnarray}
\label{RELF}
I^{(1)} \left( \frac{a_1}{a_2},\frac{a_1}{a_3} \right) = \, 
f \left( \frac{a_1}{a_2} , \frac{a_1}{a_3} \right) - \frac{1}{\epsilon_{DD}}
\, ,
\end{eqnarray}
where the anisotropy function $f$ has to be evaluated separately for the
three cases $a_1<a_2,a_3$ and $a_2<a_1,a_3$ as well as $a_3<a_1,a_2$.
\subsection{Anisotropy Function}
\begin{figure}[t]
\begin{center}
\includegraphics[scale=0.5]{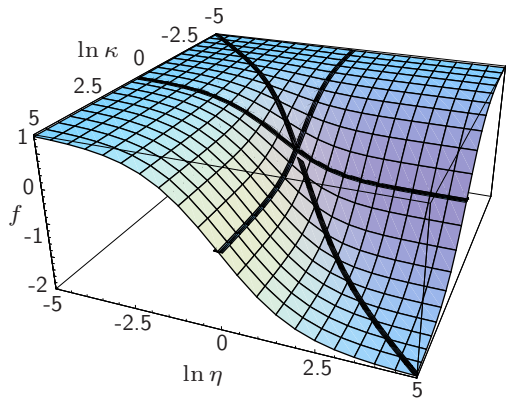}
\end{center}
\caption{\label{ANIS-f} Logarithmic plot of the anisotropy function $f(\eta,\kappa)$ in 
Eqs.~(\ref{f<<}) and (\ref{f<>}). Black curves represent the special cases
(\ref{f-SYMMETR}) with (\ref{FUNCY}). The cross line corresponds to 
$f_s ( \kappa )= f(\kappa,\kappa)$ and the lines parallel to the axes depict $f(1,\kappa)$ 
and $f(\kappa,1)$.}
\end{figure}

At first we consider the case $a_1<a_2,a_3$ and perform the 
substitution $\tau = \big( 1/a_2^2 - 1/a_3^2 \big)/z^2 - 1/a_2^2$
which results in
\begin{eqnarray}
\label{f<<}
f ( \eta, \kappa ) = 1 + \, \frac{ 3 \, \kappa \, \eta }{ \displaystyle 
\sqrt{1 \!-\! \kappa^2} \, \left( 1 \!-\! \eta^2 \right) }
\left\{ \displaystyle E \! 
\left( {\rm arcsin} \sqrt{1 \!-\! \kappa^2}, \frac{ \sqrt{1 \!-\! \eta^2} }
{ \sqrt{1 \!-\! \kappa^2} } \right) \! - F \! \left( 
{\rm arcsin} \sqrt{1 \!-\! \kappa^2}, 
\frac{ \sqrt{1 \!-\! \eta^2} }{ \sqrt{1 \!-\! \kappa^2} } \right) \right\}
\hspace*{2mm} {\rm for} \hspace*{2mm} \kappa, \eta < 1 \, .
\end{eqnarray}
For the case $a_3<a_2<a_1$ we rather use the substitution 
$\tau = \big( 1/a_1^2 - 1/a_2^2 \big)/z^2 - 1/a_3^2$, and yield
\begin{eqnarray}
\label{f<>}
f ( \eta, \kappa ) \hspace*{-1mm} &=& \hspace*{-1mm} 
\frac{1+2 \eta^2}{1-\eta^2} - \, 
\frac{ 3 \, \kappa \, \eta }{ \left( 1 \!-\! \eta^2 \right) \left( \kappa^2 
\!-\! 1 \right) } \Bigg\{ \displaystyle \sqrt{\kappa^2 \!-\! \eta^2} \, 
E \! \left( {\rm arcsin} \frac{ \sqrt{\kappa^2 \!-\! \eta^2} }{ \kappa }, 
\frac{ \sqrt{\kappa^2 \!-\! 1} }{ \sqrt{\kappa^2 \!-\! \eta^2} } \right) 
\nonumber \\*[2mm] 
&& \hspace*{-3mm} - \, 
\frac{ 1 \!-\! \eta^2 }{ \sqrt{\kappa^2 \!-\! \eta^2} } \,
F \! \left( {\rm arcsin} \, \frac{ \sqrt{\kappa^2 \!-\! \eta^2} }{ \kappa }, 
\frac{ \sqrt{\kappa^2 \!-\! 1} }{ \sqrt{\kappa^2 \!-\! \eta^2} } \right)
\Bigg\} \hspace*{3mm} {\rm for} \hspace*{2mm} 1 < \eta < \kappa \, .
\end{eqnarray}
Finally, for the case $a_2<a_3<a_1$, we have a situation which is analogous 
to the second case when performing the replacement 
$\eta \leftrightarrow \kappa$ on the right hand side of (\ref{f<>}).  
Note that all these different representations of the 
function $f$ are equivalent to each other by analytical
continuation due to the property \cite[(8.127)]{Gradshteyn}.
The importance of this anisotropy function for dipolar Bose gases
has recently been recognized in the context of expanding chromium
condensates \cite{Pedri}. There the anisotropy function appears within 
the Thomas-Fermi solution of the underlying Gross-Pitaevskii equation at 
zero temperature. Although our investigation of the critical temperature 
of dipolar gases represents a complete different physical situation, 
the anisotropy function plays here a similar important
role. Figure \ref{ANIS-f} shows how the shape of this anisotropy function 
depends on its parameters $\eta$ and $\kappa$. 
For our further investigations we have to summarize the properties of 
$f(\eta,\kappa)$, see for comparison also Ref.~\cite{Pedri}. At first, 
we mention its symmetry property
\begin{eqnarray}
f(\eta,\kappa) \, = \, f(\kappa,\eta)
\end{eqnarray}  
and, secondly, the sum rule
\begin{eqnarray}
\label{f-SUM}
f \left( \eta, \kappa \right) + f \left( \frac{\kappa}{\eta}, \frac{1}{\eta} 
\right) + f \left( \frac{1}{\kappa}, \frac{\eta}{\kappa} \right) = 0 \, .
\end{eqnarray}  
Furthermore, we need the analytical continuation of this function
into the semi-symmetric regions of the trap potential (\ref{HT})
\begin{eqnarray}
\label{f-SYMMETR}
f(\kappa,\kappa) \, = - 2 \, f(1,1/\kappa) \, \equiv \, f_s(\kappa) \, ,
\end{eqnarray}  
where the anisotropy function 
\begin{eqnarray}
\label{FUNCY}
f_s ( \kappa ) = \left\{ \hspace*{3mm}
\begin{array}{@{}ll}
{\displaystyle 
\frac{2\, \kappa^2+1}{1-\kappa^2} - 
\frac{3 \kappa^2}{(1-\kappa^2)^{3/2}}\, \mbox{artanh}\, \sqrt{1-\kappa^2}} & ,
\hspace*{0.7cm} 0 < \kappa < 1 \\*[3.5mm]
\hspace*{1.5cm} 0 & , \hspace*{1cm}\,\kappa =1 \\*[1mm]
{\displaystyle 
\frac{2\, \kappa^2+1}{1-\kappa^2} + \frac{3 \kappa^2}{(\kappa^2-1)^{3/2}}\,
\mbox{arctan}\, \sqrt{\kappa^2-1}  }  & , \hspace*{1cm}\,\kappa > 1 
\end{array} \right. 
\end{eqnarray}
describes a cylindrically symmetric dipolar BEC Refs.~\cite{Eberlein1,Eberlein2,Glaum,Pfau3}.
This function tends asymptotically to $-2$ and $1$ in the limit
$\kappa \to  \infty$ and $\kappa \to 0$, respectively (see Figure \ref{ANIS-f}).
\subsection{Results}
With the abbreviation (\ref{MI1}) we find for our perturbative 
results (\ref{FD}), (\ref{FE}) and (\ref{MSD}), (\ref{MSE}) in the three
extremal configurations $j=1,2,3$ of (\ref{ANGLES}):
\begin{eqnarray}
\label{F1AF}
{\cal F}^{({\rm D})} & = & - \frac{\mu_0 m^2}{6 \lambda^3 
(\hbar \beta \tilde{\omega})^3}
\,I^{(j)} \left( \frac{\omega_1}{\omega_2}, \frac{\omega_1}{\omega_3}
\right)\, \zeta_{\frac{3}{2},\frac{3}{2},\frac{3}{2}} \left( 
e^{\beta \mu} \right) \, ,\\
\label{F1BF}
{\cal F}^{({\rm E})} & = & -\frac{\mu_0 m^2}{6 \lambda^3 
(\hbar \beta \tilde{\omega})^3} \,
I^{(j)} \left( 1,1 \right) \, \zeta_{\frac{3}{2},\frac{3}{2},
\frac{3}{2}} \left( e^{\beta \mu} \right) \, ,\\*[1mm]
\label{F2AF}
\mu_c^{({\rm D})} & = & - \frac{\mu_0 m^2}{3 \lambda^3}\, I^{(j)}  
\left( \frac{\omega_1}{\omega_2}, \frac{\omega_1}{\omega_3} \right)\,
\zeta \left( \frac{3}{2} \right) \, ,\\*[1mm]
\label{F2BF}
\mu_c^{({\rm E})} & = &- \frac{\mu_0 m^2}{3 \lambda^3}\, 
I^{(j)} \left( 1,1 \right) \, \zeta \left( \frac{3}{2} \right) \, .
\end{eqnarray}
Here $\lambda$ denotes the thermodynamic de Broglie wave length
\begin{eqnarray}
\lambda = \sqrt{\frac{2 \pi \hbar^2 \beta}{M}} 
\end{eqnarray} 
and the generalization of the polylogarithmic function of Eq.~(\ref{FF0}) is defined 
according to
\begin{eqnarray}
\label{ABC}
\zeta_{a,b,c} (z) = \sum_{n=1}^\infty \sum_{n'=1}^\infty \frac{z^{n+n'}}
{n^a {n'}^b (n+n')^c} \, .
\end{eqnarray}
It is worth mentioning that
both exchange terms of the grand-canonical energy (\ref{F1BF}) and the 
critical chemical potential (\ref{F2BF}) depend only on the contact interaction
as the anisotropy function $f(1,1)$ vanishes
in (\ref{RELF}). Thus, within the semiclassical approach, only the direct 
contributions carrys the anisotropic character of the dipolar interaction.  
\section{Critical Temperature}\label{CT}
The total grand-canonical free energy in the three extremal configurations 
1--3 of Figure \ref{CASES} follows from (\ref{F1}) as the sum of 
(\ref{FF0}), (\ref{F1AF}), and (\ref{F1BF}):
\begin{eqnarray}
{\cal F} = - \frac{1}{\beta (\hbar \beta \tilde{\omega})^3}\,\zeta_4 
\left( e^{\beta \mu} \right) - \frac{\mu_0 m^2}{6 \lambda^3 (\hbar \beta 
\tilde{\omega})^3}\, \left[ I^{(j)}  \left( \frac{\omega_1}{\omega_2},
\frac{\omega_1}{\omega_3} \right) + I^{(j)} (1,1) \right] \,
\zeta_{\frac{3}{2},\frac{3}{2},\frac{3}{2}} \left( e^{\beta \mu} \right) 
+ \ldots \, .
\end{eqnarray}
The full problem has to be studied at fixed particle number
$N = - \partial {\cal F}/ \partial \mu$: 
\begin{eqnarray}
\label{N}
N = \frac{1}{(\hbar \beta \tilde{\omega})^3}\,\zeta_3 \left( 
e^{\beta \mu} \right) + \frac{\mu_0 m^2 \beta}{3 \lambda^3 
(\hbar \beta \tilde{\omega})^3}
\, \left[ I^{(j)}  \left( \frac{\omega_1}{\omega_2}, \frac{\omega_1}{\omega_3} 
\right) + I^{(j)} (1,1) \right] \,\zeta_{\frac{1}{2},\frac{3}{2},\frac{3}{2}} 
\left( e^{\beta \mu} \right) + \ldots \, ,
\end{eqnarray}
where we have used the identity $\displaystyle 
\zeta_{\frac{3}{2},\frac{3}{2},\frac{1}{2}}(z) = 
2 \, \zeta_{\frac{1}{2},\frac{3}{2},\frac{3}{2}}(z)$
following from (\ref{ABC}). The critical temperature $T_c$ is obtained from  
Eq.~(\ref{N}) in the limit
$\mu \uparrow \mu_c = \mu_c^{({\rm D})} + \mu_c^{({\rm E})}+\ldots$ . 
Recalling (\ref{F2AF}), (\ref{F2BF}), we obtain 
within first-order perturbation theory at the critical point 
\begin{eqnarray}
\label{N3}
N  = \, \frac{1}{(\hbar \beta_c \tilde{\omega})^3}\,\zeta (3) \, +
\frac{\mu_0 m^2 \beta_c}{3 \lambda_c^3 (\hbar \beta_c \tilde{\omega})^3} 
\left[ I^{(j)}  \left( \frac{\omega_1}{\omega_2}, \frac{\omega_1}{\omega_3} 
\right) + I^{(j)} (1,1) \right]  
\left\{ \zeta \left( \frac{1}{2},\frac{3}{2},\frac{3}{2} \right)
- \zeta \left( \frac{3}{2} \right) \zeta (2) \right\}
+ \ldots \, , 
\end{eqnarray}
where $\zeta(a,b,c)=\zeta_{a,b,c} (1)$ is a generalization of  Riemanns 
zeta-function.
The first term in Eq.~(\ref{N3}) reproduces the interaction-free particle 
number, thus leading to the critical temperature (\ref{TC0}). 
The interaction term changes this, so that the resulting shift of the critical 
temperature in configurations 1--3 reads according to (\ref{RELF}), 
(\ref{f-SYMMETR}), and (\ref{FUNCY})
\begin{eqnarray}
\label{S3}
\left( \frac{\Delta T_c}{T_c^{(0)}} \right)^{\!(1)}  &=&  - 
c_{\delta} \frac{a}{\lambda_c^{(0)}} \, + \, \frac{c_{\delta}}{2} \,  
\frac{ \mu_0 m^2 M }{ 12 \pi \hbar^2 \lambda_c^{(0)} } \, 
f \! \left( \frac{\omega_1}{\omega_2}, \frac{\omega_1}{\omega_3} \right)  \, , \nonumber \\
\left( \frac{\Delta T_c}{T_c^{(0)}} \right)^{\!(2)}  &=& - 
c_{\delta} \frac{a}{\lambda_c^{(0)}} \, + \, \frac{c_{\delta}}{2} \,  
\frac{ \mu_0 m^2 M }{ 12 \pi \hbar^2 \lambda_c^{(0)} } \, 
f \! \left( \frac{\omega_2}{\omega_3}, \frac{\omega_2}{\omega_1} \right) \, ,
\nonumber \\
\left( \frac{\Delta T_c}{T_c^{(0)}} \right)^{\!(3)} & = & - 
c_{\delta} \frac{a}{\lambda_c^{(0)}} \, + \, \frac{c_{\delta}}{2} \,  
\frac{ \mu_0 m^2 M }{ 12 \pi \hbar^2 \lambda_c^{(0)} } \, 
f \! \left( \frac{\omega_3}{\omega_1}, \frac{\omega_3}{\omega_2} \right) 
\, .
\end{eqnarray} 
The dimensionless prefactor $c_{\delta}$ for the contact interaction has 
the value
\begin{eqnarray}
\label{CDelta}
c_\delta = \frac{4}{3 \zeta(3)}\,\left[ \zeta \left( \frac{3}{2} \right) \,
\zeta (2) - \zeta \left( \frac{1}{2},\frac{3}{2},\frac{3}{2} \right)
\right] \approx 3.426 \, .
\end{eqnarray}
Now we discuss the physical implications of our first-order perturbative 
result (\ref{S3}). By setting $m=0$ we obtain for all configurations 
1--3 in Figure~\ref{CASES} the same downwards shift of the 
critical temperature with the dimensionless prefactor (\ref{CDelta}). This 
simple result for the isotropic contact interaction was originally
derived in Ref.~\cite{Giorgini} within a mean-field approach. Only recently 
it was confirmed experimentally by investigating the onset of Bose-Einstein 
condensation in the 
hyperfine ground state of ${}^{87}$Rb \cite{Aspect}. 
Furthermore, our result (\ref{S3}) also shows
how the shift of the critical 
temperature depends on the anisotropic magnetic dipole-dipole interaction.\\

So far we restricted our detailed analysis to those spherical angles 
$(\alpha,\phi)$ between the symmetry axes of the interaction (\ref{MINT}) and 
the trap (\ref{HT}) which coincide with the extremal cases (\ref{ANGLES}). 
It is straight-forward to extend our result to arbitrary angles. 
For this purpose we use the master integral (\ref{MIint}) and its
decomposition (\ref{MI-II-III}) into the integral (\ref{MI1}) for $j=1,2,3$.
With this we calculate 
the following angle-dependent shift of the critical temperature:
\begin{eqnarray}
\label{S-alfa}
\left( \frac{\Delta T_c}{T_c^{(0)}} \right)^{\!(\alpha, \phi)} \!\!\!=
- c_{\delta} \, \frac{a}{\lambda_c^{(0)}} \, + \, \frac{c_{\delta}}{2} \, 
\frac{ \mu_0 m^2 M }{ 12 \pi \hbar^2 \lambda_c^{(0)} } \, 
\Bigg\{ \! \sin^2 \! \alpha \bigg[ \cos^2 \! \phi \, 
f \! \left( \frac{\omega_1}{\omega_2}, \frac{\omega_1}{\omega_3} \right) 
+ \sin^2 \! \phi \, 
f \! \left( \frac{\omega_2}{\omega_3}, \frac{\omega_2}{\omega_1} \right)
\! \bigg] \!
+ \cos^2 \! \alpha \,
f \! \left( \frac{\omega_3}{\omega_1}, \frac{\omega_3}{\omega_2} \right) 
\!\! \Bigg\} .
\end{eqnarray} 
For the special cases of $\alpha=0, \alpha=\pi/2$ and 
$\phi=0, \pi/2, \pi, 3\pi/2$, Eq.~(\ref{S-alfa}) reduces to the previous 
results (\ref{S3}), which turn out to be extremal. 
\section{Magic Angles}\label{MANG}
Changing the angles $(\alpha,\phi)$ between the symmetry axes of the trap
and the magnetization direction allows us to tune the dipolar effect between 
those maximal and minimal values. In particular, there exist so-called magic angles, where
the dipolar effects vanish.
We read off from Eq.~(\ref{S-alfa}) that those angles are given for an 
arbitrary anisotropy due to the sum rule (\ref{f-SUM}) by
\begin{eqnarray} 
\label{MAGIC}
\alpha_0 (\phi) = {\rm arccot} \, \Bigg\{ \! \pm \sqrt{ \, 
\sin^2 \! \phi - \cos \! 2 \phi \,\, f \! \left( \frac{\omega_1}{\omega_2}, 
\frac{\omega_1}{\omega_3} \right) \! \bigg/ f \! \left( 
\frac{\omega_3}{\omega_1}, \frac{\omega_3}{\omega_2} \right)
} \, \Bigg\} \, .
\end{eqnarray} 
Note that the radicand has to be positive, which restricts the range of values for the
polar angles $\phi$. In general,
the possibility to find a magic angle depends crucially on the
anisotropy coefficients. Only in the case that $f ( \omega_3 / \omega_1 , 
\omega_3 / \omega_2 ) < - f ( \omega_1 / \omega_2 ,
\omega_1 / \omega_3 )$ is fulfilled, magic angles always exist.\\

\begin{figure}[t]
\begin{center}
a)\includegraphics[scale=0.45]{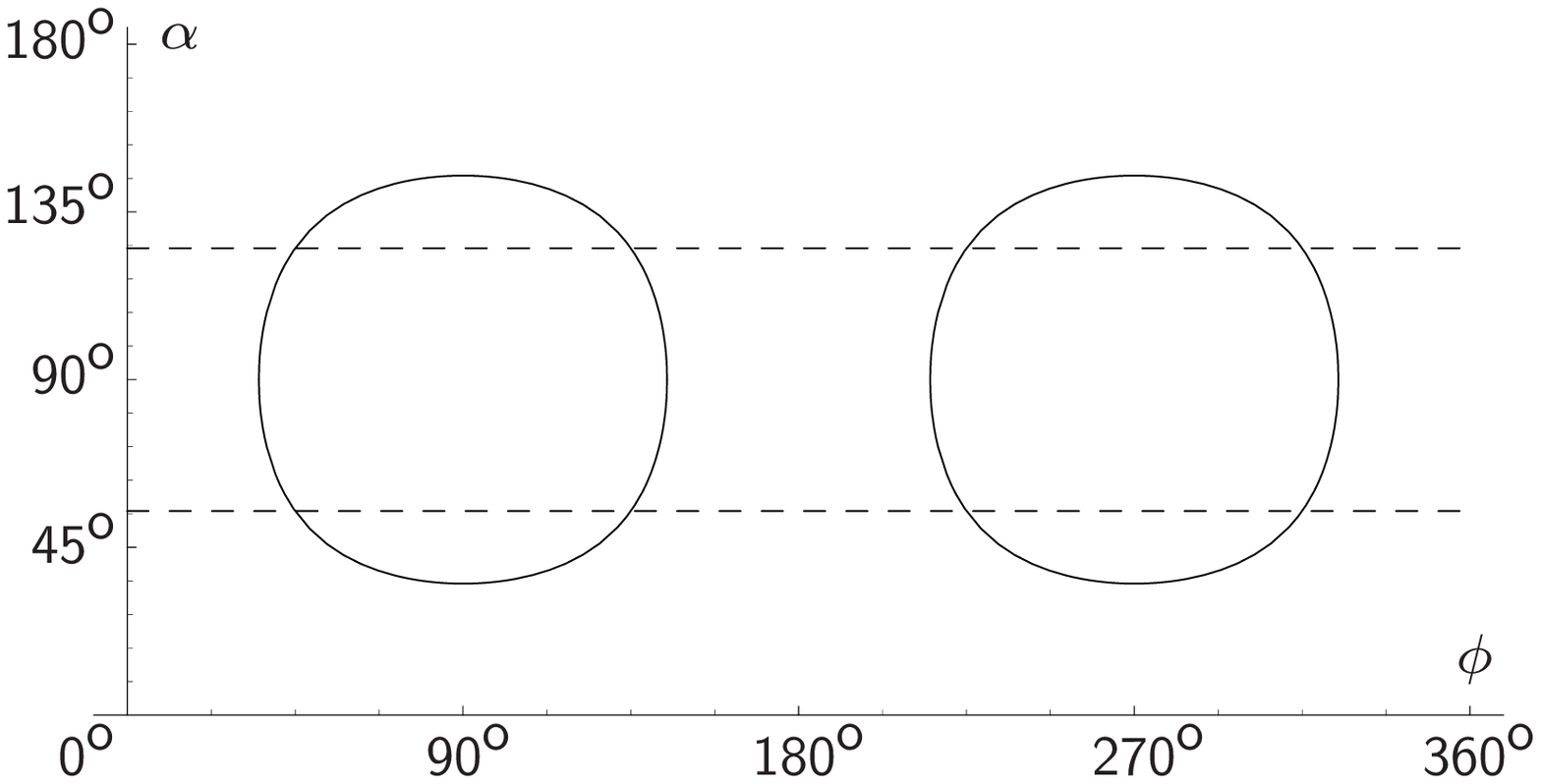}
\hspace*{1cm}
b)\includegraphics[scale=0.45]{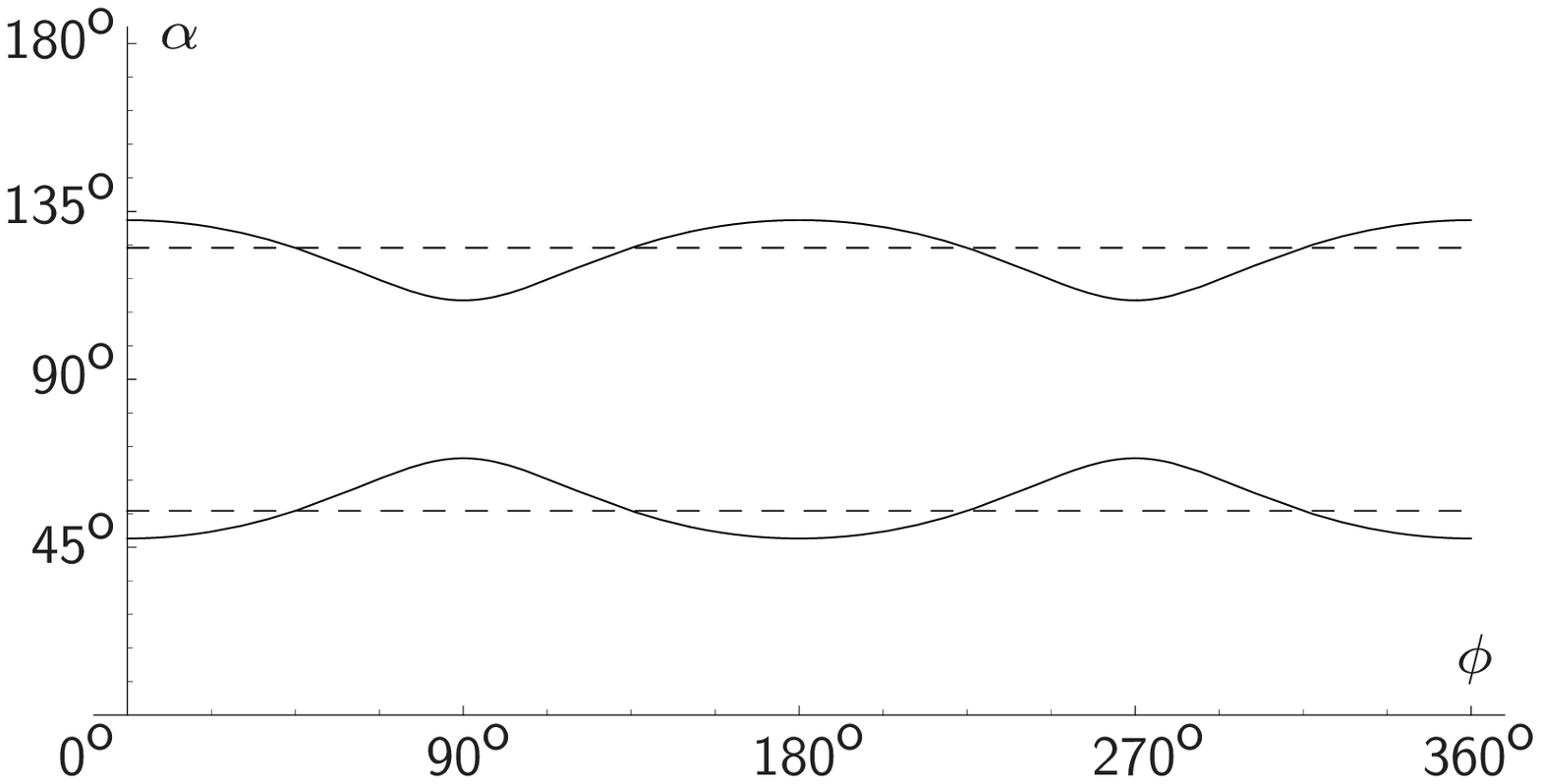}
\end{center}
\caption{\label{magic-fi} 
The location of magic angles $(\alpha,\phi)$, where the dipolar effect
vanishes. a) The solid line stands for a cylinder symmetric
problem with $\omega_1 = \omega_3$ and the dashed line for that with
symmetry $\omega_1 = \omega_2$. b) The solid line stands for the anisotropic 
parameters of the Stuttgart experiment 
and the dashed line for a cylinder symmetric problem with $\omega_1 = \omega_2$.}
\end{figure}

Now we present the results for cylindrically symmetric traps. 
At first, we mention for $\omega_1=\omega_2$ that the 
magic angle (\ref{MAGIC}) reduces to a value, which is independent of the 
second anisotropy parameter $\omega_1/\omega_3$. This value is  also 
$\phi$-independent according to (\ref{MAGIC}) with (\ref{f-SYMMETR}) and amounts to 
$\alpha_0 = {\rm arccot} ( \pm 1/ \sqrt{2} ) = 54.7^{\circ}$ or 
$125.3^{\circ}$ \cite{Pfau2,Glaum}. 
For another symmetric case with $\omega_1=\omega_3$ we find with 
(\ref{f-SYMMETR}) that
$f ( \omega_3 / \omega_1 , \omega_3 / \omega_2 ) = f ( \omega_1 / \omega_2 ,
\omega_1 / \omega_3 )$, thus solutions of (\ref{MAGIC}) only exist for $\sin^2 \phi > 1/3$.
The locations of magic spherical angles for both
symmetric cases $\omega_1=\omega_2$ and $\omega_1=\omega_3$ are depicted 
in Figure \ref{magic-fi} a). For the third symmetric case $\omega_2=\omega_3$ the 
situation is similar to $\omega_1=\omega_3$ as the
polar angle $\phi$ is only shifted by $\pi / 2$.\\

Furthermore,
we discuss the situation in a complete anisotropic harmonic trap by the example of
the trap frequencies of the Stuttgart experiment \cite{Pfau6} which
are given by $\omega_1 = 2 \pi \cdot 581\,\mbox{Hz}$, 
$\omega_2 = 2 \pi \cdot 406\,\mbox{Hz}$, and
$\omega_3 = 2 \pi \cdot 138\,\mbox{Hz}$.
Thus, we find from (\ref{f<>}) the values
$f ( \omega_1 / \omega_2 , \omega_1 / \omega_3 )= -0.6252$ and 
$f ( \omega_3 / \omega_1 , \omega_3 / \omega_2 )= +0.7356$. 
Specializing the general result (\ref{MAGIC}) for the Stuttgart experiment leads to
\begin{eqnarray} 
\label{EXPMAG}
\alpha_0 (\phi) = {\rm arccot} \left\{  \pm \sqrt{ \, 0.150
\sin^2 \! \phi + 0.850 \cos^2 \! \phi } \,\, \right\} \, .
\end{eqnarray} 
This behaviour is depicted in
Figure \ref{magic-fi} b), where we see the location of the magic angles 
compared with the $\phi$-independent value of a problem with a cylinder
symmetry around the $z$-axis. \\
\section{Stuttgart Experiment}\label{SEES}
Here we discuss the consequences of our results for the concrete example of
the parameters from the ongoing experiments on the Bose-Einstein
condensation of ${}^{52}$Cr-atoms at the University of Stuttgart \cite{Pfau6}. 
There the geometric 
mean frequency is $\tilde{\omega}=2\pi \cdot 319\,\mbox{Hz}$.
The total atom number is $N=100\,000$, yielding an interaction-free critical 
temperature (\ref{TC0}) of about $T_c^{(0)} = 670 \, \mbox{nK}$. 
The finite-size correction of $T_c^{(0)}$ in the Stuttgart experiment 
follows from (\ref{FS}) and amounts 
to $-1.8$ \%. This can now be compared with a shift of the critical temperature
due to the contact and the magnetic dipole-dipole 
interaction.\\

\begin{figure}[t]
\begin{center}
a)\hspace*{-5mm} \includegraphics[scale=0.6]{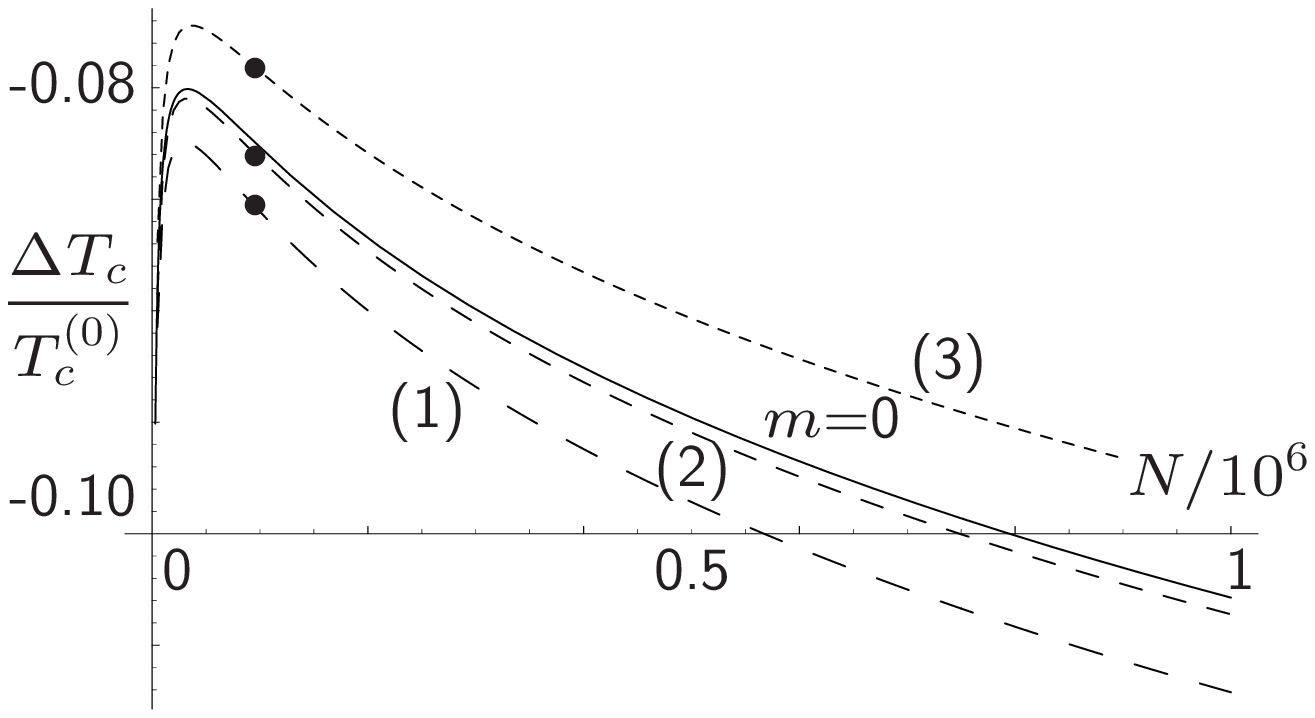}
\hspace*{1cm}
b)\hspace*{-2mm}\includegraphics[scale=0.53]{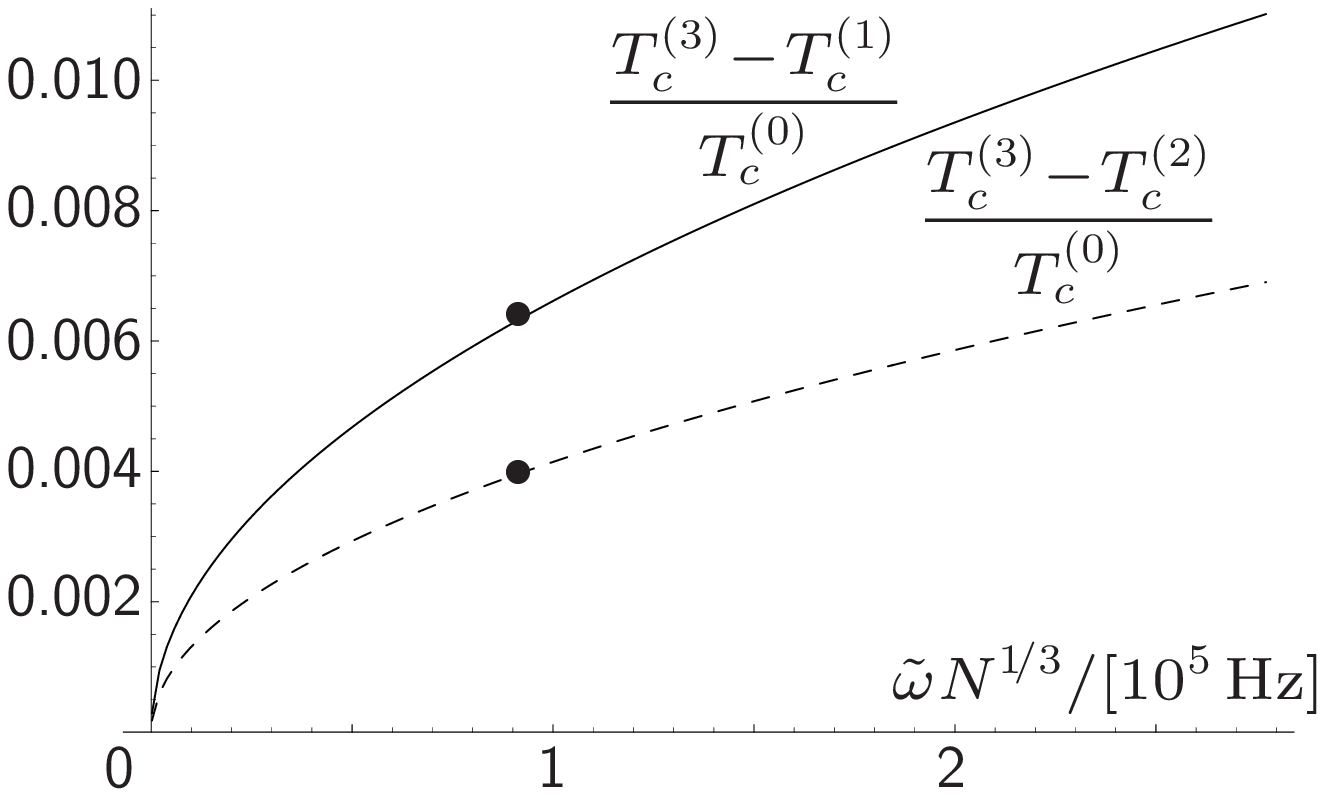}
\end{center}
\caption{\label{RES}a) Shift of the critical temperature $\Delta T_c$ with 
respect to the interaction-free critical temperature $T_c^{(0)}$ for a 
${}^{52}$Cr gas in a harmonic trap with frequencies 
$\omega_1 = 2 \pi \cdot 581\,\mbox{Hz}$, 
$\omega_2 = 2 \pi \cdot 406\,\mbox{Hz}$, and 
$\omega_3 = 2 \pi \cdot 138\,\mbox{Hz}$ calculated from (\ref{S3}) and 
(\ref{FS}) without (straight line) and 
with (dashed lines) magnetic
dipole-dipole interaction for the  cases 1--3 of Figure~\ref{CASES}.
b) Differences of the critical temperature shifts versus effective 
parameter $\tilde{\omega} N^{1/3}$. The respective dots
indicate the present parameters of the Stuttgart ${}^{52}$Cr experiment.}
\end{figure}

The scattering length of the ${}^{52}$Cr-atoms is given by 
$a \approx 105\,a_B$ with the Bohr radius $a_B$ \cite{Pfau8}, so we obtain from (\ref{EDD})
$\epsilon_{DD}=0.144$. The 
thermodynamic de Broglie wave length  has the value 
$\lambda_c^{(0)} \approx 5\,598\,a_B$ for the parameter of Stuttgart 
experiment. From Eq.~(\ref{S3}) we 
obtain from the contact interaction a downwards
shift of the critical temperature by $-6.4$ \%. This is now modified by the 
magnetic dipole-dipole interaction. 
At first, we consider case 1 from Figure \ref{CASES}, in which the 
magnetization is parallel to the trap axes with the largest moment of inertia.
There we find from (\ref{f<>}) the result 
$f ( \omega_1 / \omega_2 , \omega_1 / \omega_3 )= -0.6252$,  
which leads via Eq.~(\ref{S3}) to an extra downward shift of the critical 
temperature by $-0.29$ \% due to the magnetic dipole-dipole interaction. 
For the case 2 of Figure \ref{CASES} we find 
$f ( \omega_2 / \omega_3 , \omega_2 / \omega_1 )= -0.1104$,
which amounts to an additional downward shift of the critical 
temperature due to the magnetic dipole-dipole interaction by $-0.05$ \%.
Finally, for case 3 of Figure \ref{CASES} with the magnetization direction 
parallel to the trap axes with the smallest moment of inertia, we find
$f ( \omega_3 / \omega_1 , \omega_3 / \omega_2 )= +0.7356$ which 
results in an upward shift of the critical 
temperature due to the dipole-dipole interaction of about $+0.34$ \%.
Figure \ref{RES}a) shows the resulting total shift of the
critical temperature $\Delta T_c$ for the ${}^{52}$Cr gas with respect to the
interaction-free critical temperature $T_c^{(0)}$ versus the particle number 
$N$. Both the finite-size corrections and the contact interaction lead to 
a huge shift, on top of which the small dipolar effect is seen. \\

\begin{figure}[t]
\begin{center}
a)\hspace*{-3mm}\includegraphics[scale=0.37]{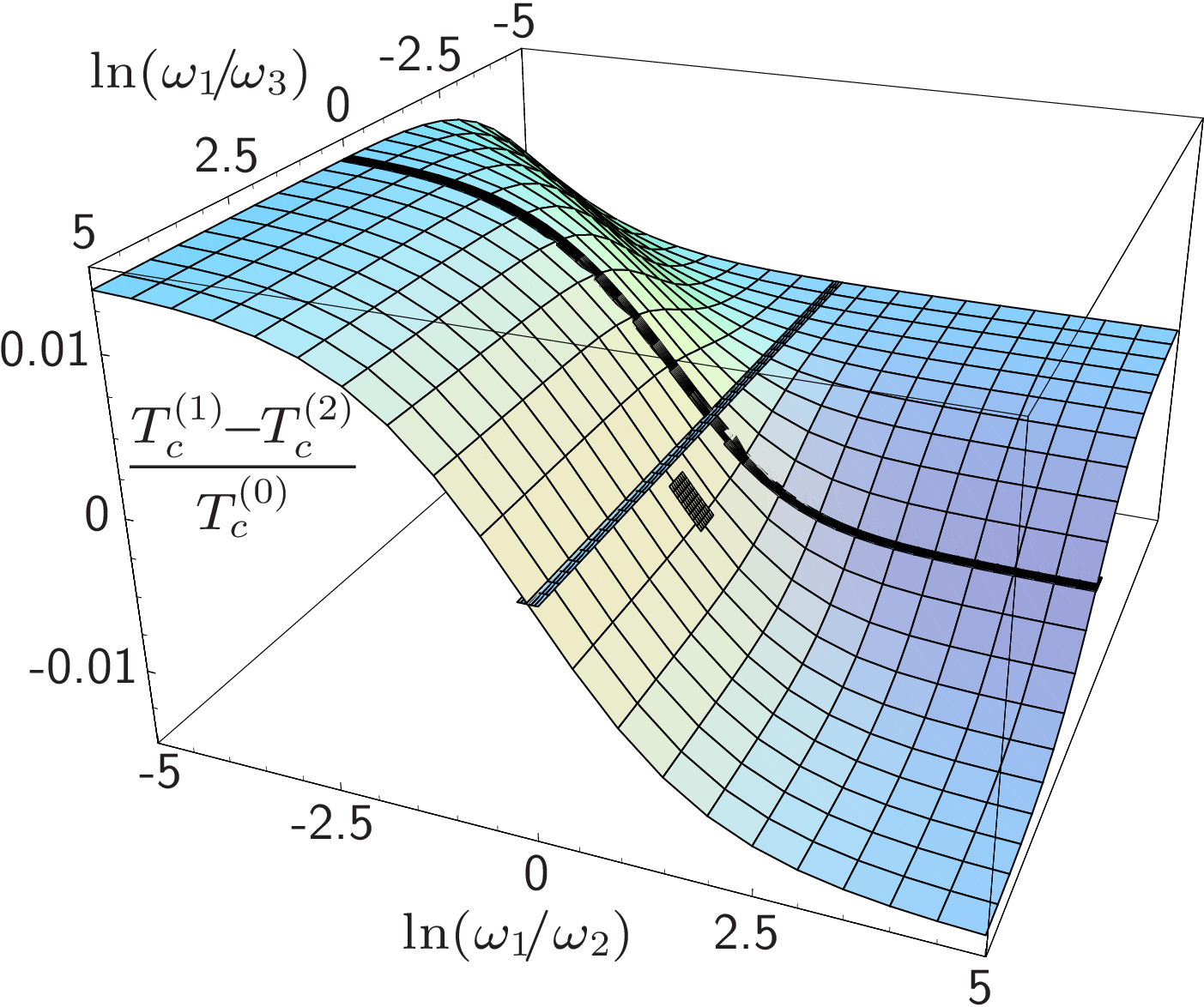}
\hspace*{0.5cm}
b)\hspace*{-3mm}\includegraphics[scale=0.37]{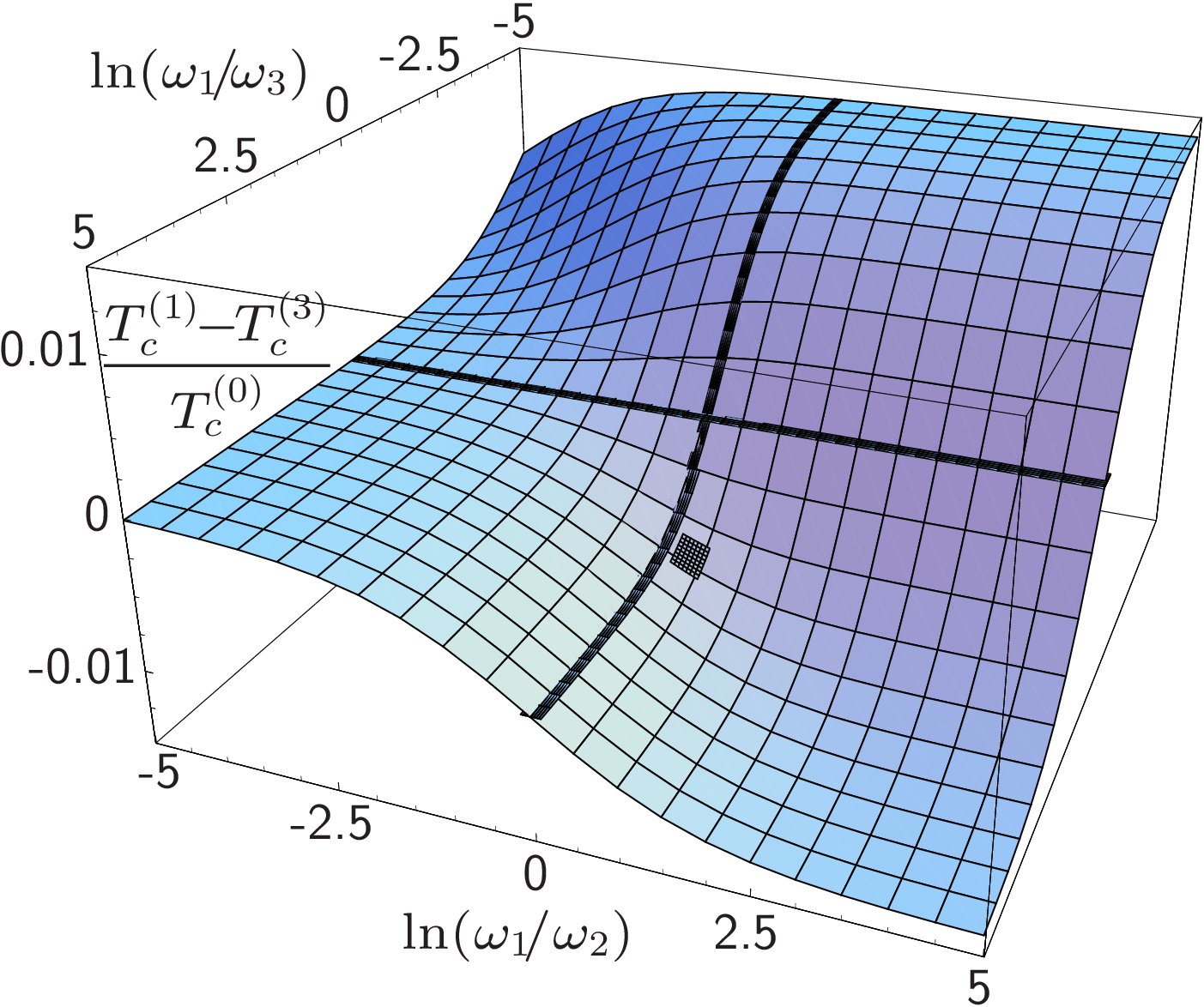}
\hspace*{0.5cm}
c)\hspace*{-3mm}\includegraphics[scale=0.37]{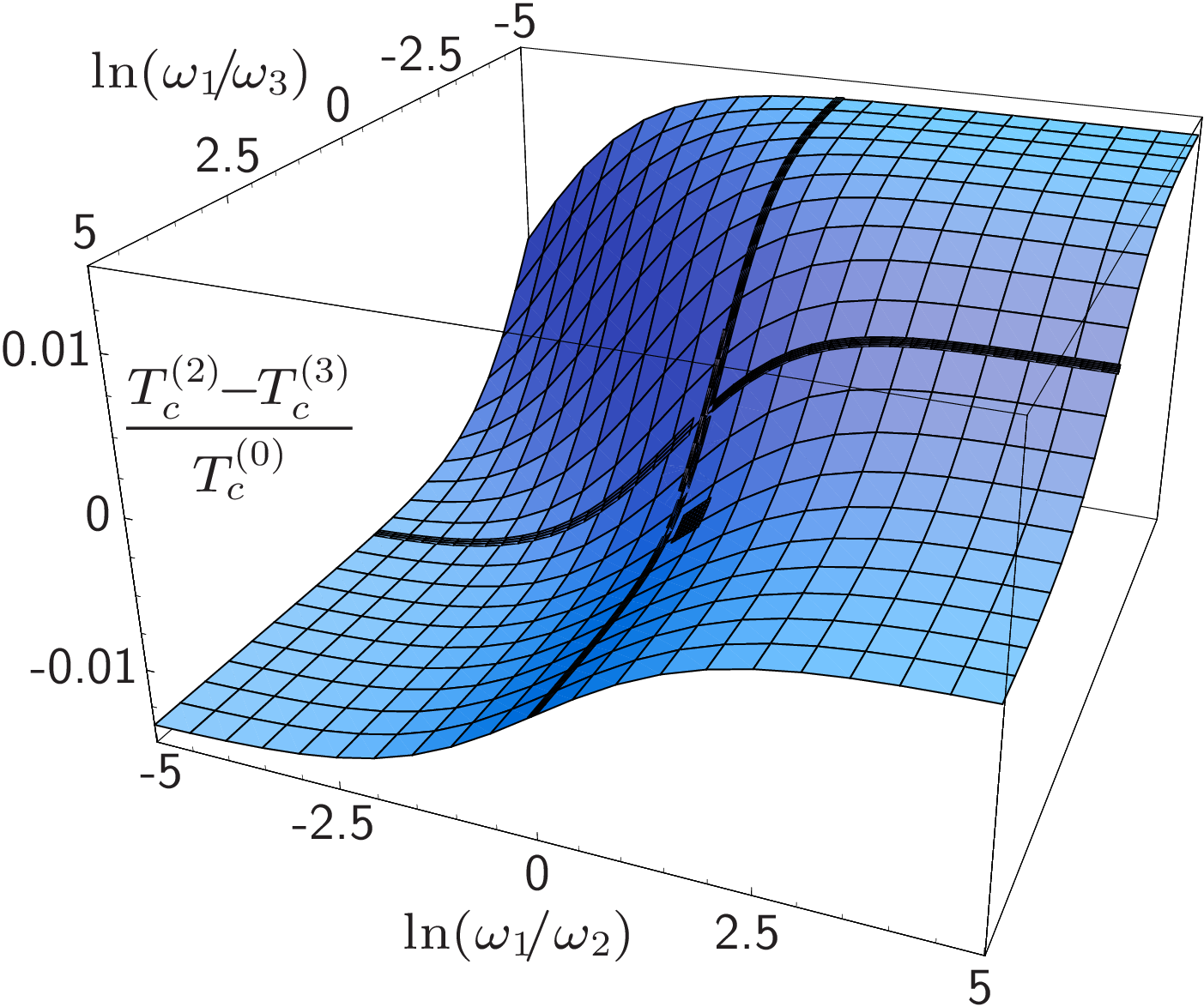}
\end{center}
\caption{\label{RES-kappa} Shifts of critical temperatures a) 
$\Delta T_c^{(1)} \!-\! \Delta T_c^{(2)}$, 
b) $\Delta T_c^{(1)} \!-\! \Delta T_c^{(3)}$
and c) $\Delta T_c^{(2)} \!-\! \Delta T_c^{(3)}$ 
with respect to the interaction-free critical temperature $T_c^{(0)}$ for 
$N \!=\! 10^5$ ${}^{52}$Cr-atoms in a trap with mean frequency 
$\tilde{\omega}= 2 \pi \cdot 319$ Hz versus anisotropy parameters 
$\omega_1/\omega_2$ and $\omega_1/\omega_3$ according to (\ref{S3}). The black
curves represent cylindrically symmetric cases and the black squares the 
values of the Stuttgart experiment.}
\end{figure}

Furthermore, we we follow Ref.~\cite{Glaum}  and
suggest to measure differences between the critical 
temperatures in the three extremal cases of Figure \ref{CASES}. 
The large shift caused by the finite-size corrections (\ref{FS}) and the 
contact interaction would cancel due to their isotropic character. 
Thus, the difference between the shifts is exclusively caused by the 
magnetic dipole-dipole interaction. For the total atom number $N=100\,000$
and the mean frequency $\tilde{\omega}=2 \pi \cdot 319\,\mbox{Hz}$, 
the temperature difference between the case 1 and 3 amounts to a net effect of 
0.61 \% of the interaction-free critical temperature $T_c^{(0)}$.
This dipolar effect increases with the geometric mean frequency 
$\tilde{\omega}$ and the total atom number $N$ as 
$[ \tilde{\omega} N^{1/3} ]^{1/2}$, as is shown in Figure~\ref{RES}b). 
Another possibility to enhance the difference of the critical
temperatures is provided by varying the anisotropy strengths 
$\omega_1/\omega_2$ and $\omega_1/\omega_3$ of the harmonic trap potential
as seen in Figure \ref{RES-kappa}. For instance, in order 
to increase the difference of the critical temperature between the first and 
third configuration, we need to increase the aspect ratio $\omega_1/\omega_2$.
Furthermore, it is preferable to work with moderate anisotropy parameters 
$\omega_1/\omega_3 <1$ (pancake like) than in the regime 
$\omega_1/\omega_3 > 1$ (cigar shaped) of the Stuttgart experiment, as is 
indicated by Figure \ref{RES-kappa} b).\\

Finally, we elucidate how to achieve measurable dipolar effects at the critical 
point. At first, varying the anisotropy parameters only allows to 
increase the critical 
temperature difference $T_c^{(3)} \!-\! T_c^{(1)}$ with respect to the 
interaction-free value from currently 0.61 \% up to a maximal value of around
1.39 \%. Note that the latter represents an upper limit according to our 
perturbative calculations. Not so strictly limited is the 
possibility to amplify the magnetic dipolar effect by 
increasing the particle number or the mean trap frequency. For instance, a 
doubling of the latter yields by constant particle number an increase of 
the effect by the factor 1.41. An even higher increase of the particle number
seems to be promising, but is in fact physically limited by three-body 
losses. 
\section{Conclusion}
So far, the dipolar nature of nowdays available BECs has only been resolved in 
expansion experiments \cite{Pfau7}. The analysis of the present article 
shows that it could become possible to detect a signal of the underlying 
dipole-dipole interaction also from measuring the critical 
temperature of the onset of Bose-Einstein condensation.
Most sensitive is the difference of the critical temperatures 
where the magnetization direction is parallel to the  
axes of the harmonic trap with largest and smallest moment of inertia. 
We have shown quantitatively that this temperature difference increases with
the number of chromium atoms, the geometrical mean frequency, and the 
anisotropy of the trap. Furthermore, our results will, certainly, be useful also for 
other dipolar systems with a tunable
dipole moment, like heteronuclear molecules in low vibrational states 
\cite{Weinstein,Doyle,Meijer,Sage,Wang,Mancini,Sengstock}, where the dipolar effect will
be larger.\\
\section*{Acknowledgement}
It is a pleasure to thank Hagen Kleinert, Tilman Pfau, and Shai Ronen for stimulating 
discussions. Furthermore, we acknowledge financial support from the DFG
Priority Program SPP 1116 {\it Interactions in Ultra-Cold Atomic and 
Molecular Gases}. 
%

%
\end{document}